# Financial bubbles: mechanisms and diagnostics

**Didier Sornette and Peter Cauwels**
*ETH Zurich*
*Chair of Entrepreneurial Risk*
*Department of Management Technology and Economics*

8 April 2014

**Abstract:** Financial bubbles are subject to debate and controversy. However, they are not well understood and are hardly ever characterised specifically, especially ex ante. We define a bubble as a period of unsustainable growth, when the price of an asset increases ever more quickly, in a series of accelerating phases of corrections and rebounds. More technically, during a bubble phase, the price follows a faster-than-exponential power law growth process, often accompanied by log-periodic oscillations. This dynamic ends abruptly in a change of regime that may be a crash or a substantial correction. Because they leave such specific traces, bubbles may be recognised in advance: that is, before they burst. In this paper, we will explain the mechanism behind financial bubbles in an intuitive way. We will show how the log-periodic power law emerges spontaneously from the complex system that financial markets are as a consequence of feedback mechanisms, hierarchical structure and specific trading dynamics and investment styles.

We argue that the risk of a major correction, or even a crash, becomes substantial when a bubble develops towards maturity, and that it is therefore very important to find evidence of bubbles and to follow their development from as early a stage as possible. The tools that are explained in this paper actually serve that purpose. They are at the core of the Financial Crisis Observatory at the ETH in Zurich where tens of thousands of assets are monitored on a daily basis. This allows us to have a continuous overview of emerging bubbles in the global financial markets. The report available as part of the Notenstein white paper series (2014) with the title "Financial Bubbles: Mechanism, diagnostic and state of the world (Feb. 2014)" presents a practical application of the methodology outlined in this article and describes our view on the status of positive and negative bubbles in the financial markets, as of the end of January 2014.

**Keywords:** systemic crisis, change of regime, financial bubble, bifurcation, instability, precursors, singularity, prediction, super-exponential, log-periodic, hierarchies, positive feedbacks, procyclicality,

**Contents**

## 1- The dog that did not bark

In a tale by Arthur Conan Doyle, Sherlock Holmes draws the attention of Scotland Yard detectives to the curious behavior of the dog in the night when the race horse Silver Blaze was stolen. One of the officers objects that the dog did nothing. "That, was the curious incident", replies Holmes. While the inspectors are totally absorbed by the nitty-gritty details of the case, Holmes takes a step back to get a broader perspective. Thanks to this eccentric approach, he quickly finds the key to solve the mystery. The horse's trainer committed the crime. For once, it was not the butler.

This story is reminiscent of what Paul Krugman, the winner of the 2008 Nobel Memorial Prize in Economic Sciences, calls *up-and-down economics*. Individual market movements are scrutinized as if they provide crucial clues to solve a mystery. Especially in the media and on the blogosphere, up-and-down price movements are covered like the exchange of shots during a tennis match. One day, stocks are down, fear is switched on and the Euro zone is in danger of total collapse. The next day, the whole lot turns around and Europeans can heave a sigh in relief. In this soap opera of self-fulfilling prophecy and wishful thinking, it is difficult to find the forest through the trees and to focus on the real nature of the financial system hidden behind the ephemeral details.

By its nature, trading is carried out by interacting players connected in a hierarchical network structure who affect one another continuously. As such, financial markets are open, adaptive, out-of-equilibrium systems that are subject to nonlinear dynamics, created particularly but not only by imitation and herd behaviour. Such structures are extremely hard to model, and it is practically impossible to understand their behaviour in full detail. In fact, this assertion can be formalised by algorithmic Information theory – a combination of Shannon's information theory and Turing's computability theory – stating that, in a nutshell, most dynamic systems are intrinsically unpredictable ("computationally irreducible"). This is related to Gödel's incompleteness theorem in mathematical logic.

However, if we take a step back and look at the broader, lower-resolution picture, predictability creeps back in and one can often be surprised to discover unexpected emerging macroscopic properties, like financial bubbles. The realisation that predictability is actually possible, if the right scales are chosen and thoughtful questions are asked, is a key insight on which our research in the last two decades has been built.

This can be illustrated by many everyday-life observations. When the plug is pulled out of the bathtub it opens the system and pushes it out of equilibrium; a large-scale vortex will then often appear. A detailed knowledge of the water molecules, of the bathtub's shape, or of the force used to pull the plug out will not help to understand the emergence of this coherent structure. Other fascinating examples of emergent macroscopic organisations are schools of anchovies or flocks of starlings, capable, in the absence of any conductor or leader, of beautiful choreographic movements, which emerge from the repeated use of basic rules of repulsion and attraction between the individual animals. Similarly, the heart beats as a result of self-organised rotating spiral waves of electro-mechanical activity that contribute to a spontaneous pacemaker. Out of the apparent chaos at a lower level emerges a predictable order at a higher level.

To understand the complexity around us, and especially in economics and finance, we need to step back from traditional concepts such as utility maximisation or equilibrium in closed systems, and use innovative ways of thinking to discover sometimes surprisingly simple rules. This is our line of attack. We look at financial markets from an alternative viewpoint, in the way Sherlock Holmes solves mysteries. This allows us to better understand the specific emerging macroscopic structure of financial bubbles, the extraordinary occurrence of great over-valuation over transient periods that, we argue, largely shapes the way economies and nations develop. Like swarms of birds or fishes, bubbles are the result of subtle forms of interaction and organisation within a system – this time, the financial system.

In the following section, we give an intuitive explanation of our bubble-tracking model. We argue that the risk of a major correction, or even a crash, becomes substantial when a bubble develops towards maturity, and that it is therefore very important to find evidence of bubbles and to follow their development from as early a stage as possible. This is the goal of the Financial Crisis Observatory (FCO) at the ETH Zurich (the Swiss Federal Institute of Technology), where we monitor more than twenty-five thousand financial time series on a daily basis.

First, though, let us explain why bubbles appear, how they are intimately linked to our financial system, and why studying and understanding them has never been more relevant than today.

## 2- Economics of boom and bust

In an article for the Notenstein Academy White Paper Series entitled *A Creepy World* [1], we argued that the dynamics behind business cycles are highly nonlinear and extremely asymmetric. We compared this with mechanical creep, where long periods of falsely perceived stability and status quo are terminated by extreme events that reset the system. In business, this is caused by the dynamic interaction between innovation and speculation. As a wave of new innovations flourishes, a boom sets in. New companies are created with the purpose of producing and commercialising these new technological inventions. These companies, however, are generally cash-poor. The wave of innovation can proceed only through financial intermediation from established, cash-generating enterprises to new, cash-absorbing businesses. As a result, periods of high innovation are accompanied by an increase in the size of the financial industry. William Janeway, an American venture capital investor, with over 40 years of experience, has provided an accessible way of understanding the dynamics behind the innovation economy in his recent book [2]. In his view, informed both by practice and theory, the innovation economy begins with discovery and culminates in speculation, with continuous positive feedback loops between the two. Over some 250 years, so his argument goes, economic growth has been driven by successive processes of trial and error: upstream explorations in research and inventions and downstream experiments in exploiting the new economic space opened by innovation. This interpretation offers an original theory about the role of asset bubbles in financing technological innovation, and about the specific role played by the state in enabling innovation processes [2, 3]. The over-reach associated with the bubbles and the disproportional growth of the banking industry finally causes a collapse and puts an end to the cycle. The process then can start again with a new wave of innovations.

In "*1980-2014, the Illusion of the Perpetual Money Machine and what it bodes for the future*" [4], we explained that we are currently in such an over-reach phase of disproportionate financialisation, associated with a succession of bubbles and crashes culminating in the financial crisis of 2008. This prompted central banks worldwide to take unprecedented actions, including an extended and still on-going period of very low interest rates, as well as "unconventional" policies such as asset purchases and long-term provision of liquidity to banks. This is what the International Monetary Fund (IMF) refers to as "Monetary Policy Plus" (MP-plus). In a recent Global Financial Stability Report [5], an explicit warning is given: *"… MP-plus may have undesirable side effects, including some that may put financial stability at risk. Ample bank liquidity may raise credit risk at banks by compromising underwriting and loan quality standards, and it may encourage a delay in necessary balance sheet repair and bank restructuring. Likewise, low interest rates encourage other financial institutions, including pension funds, insurance companies, and money market mutual funds, to increase risk by "searching for yield". A search for yield can help push the market value of some assets beyond their fundamental value ("bubbles") …"*

To fight the crisis that was the most recent culmination of a recurring pattern of bubbles and crashes, central banks introduced MP-plus, about which the IMF warns that it increases the risk of … bubbles. This anomaly is well captured by the shocking statement of L. Summers, previous Secretary of the US Treasury and president emeritus of Harvard University, in an article in the Financial Times on 12 June 2011: "*The central irony of the financial crisis is that while it is caused by too much confidence, borrowing and lending, and spending, it is only resolved by increases in confidence, borrowing and lending, and spending*."

In these uncertain times, a better understanding of financial bubbles, which are hardly ever characterised specifically, has never been more relevant.

**3- Bubble mechanics**

*3.1 Fundamental value, efficient markets and irrational exuberance*
During bubbles, prices move away from their so-called fundamental value; where, during positive bubbles, there is excessive demand and, during negative bubbles, there is disproportionate selling. But, before going deeper into explaining and defining bubbles, what do we actually know about this "fundamental value"?

In his 1985 presidential address to the American Finance Association [6], Fischer Black elaborated on the effect of noise in financial markets. He recognised that, without noise, there would be no trading, but the reverse of the medal is that, because of noise, "*We are forced to act largely in the dark*" [6]. An efficient market, according to Black, "*is one in which price is within a factor of two of (the true) value*" [6].

Black's heuristic can be easily checked using the Gordon-Shapiro dividend discount model. The formula (given in Equation 2 below) is highly intuitive, and is based on the fact that the total return *r* obtained by investing in a stock is the sum of its dividend yield *D/P* (where *D* is the dividend and *P* is the price) and of the growth rate *g* of the dividend (which is also the growth rate of the price):

$$r = \frac{D}{P} + g \qquad (1)$$

Rearranging the factors gives us the basic stock pricing formula:

$$P = \frac{D}{r - g} \qquad (2)$$

Suppose that, on an annual basis, the expected dividend (D) is 100, the total return (r) 8 percent and the growth rate (g) 4 percent. Then, the stock would be worth 2,500. Now, by simply reducing the expected total return from 8 percent to 6 percent, the expected stock price doubles to 5,000, for the same dividend and growth expectations. Reducing the expected return increases the price: this may seem counter-intuitive but simply results from the fact that *r* is also the rate by which to discount the future dividends; thus a smaller discount rate gives more value to future incomes and so increases the present price. The same mechanism links yields and bond prices. In the long run, adjusted for annual inflation, US stock returns have averaged about 7 percent. So, if one investor uses an expectation of 6 percent and the other one of 8 percent, the difference between their pricing will be a factor of two.

Clearly, this is a simplification, and other more advanced discounted cash flow models exist. Nevertheless, the conclusion generally holds: when pricing stocks, estimates of different factors such as dividends, returns and growth expectations are needed, small differences in these estimates, especially those appearing in the denominator, may result in large deviations in the expected price. As most estimates are based on historical data, valuing stocks is a bit like driving in the dark while looking in the rear-view mirror, even in efficient markets.

This reasoning would suggest that the search for the fair value of an asset or a project is a vain quest. However, this is where the power of the financial markets comes into play: by voting with his or her wallet, each investor contributes to the collective knowledge, and the aggregate information is revealed in the asset price. This mechanism driven by "the wisdom of crowds" transcends the above formula and provides prices that are in general reasonable and reflect a useful and workable consensus. In financial jargon, this is called the "efficient-market hypothesis", introduced by Paul Samuelson and Eugene Fama, the latter being the joint recipient of the 2013 Nobel Memorial Prize in Economic Sciences.

Robert Shiller, one of the other recipients of the same prize, was honoured ironically for spearheading the opposite notion that markets, at times, exhibit "irrational exuberance" [7]. While the efficient-market hypothesis provides a useful first-order representation of financial markets in normal times, situations can be observed in which the anchor of a fundamental price is shaky and future gains are characterised by serious uncertainties. Our research shows that the absence of a strong anchor, a well-defined fundamental value to which the price can be attached, provides a fertile environment for the occurrence of bubbles. When a number of additional elements are present, markets experience transient phases in which they disconnect in specific, dangerous ways from the fundamental value. These are situations in which investors are following the herd, and pushing a price up on an unsustainable growth trajectory. We define these as financial bubbles and explain their specific characteristics in the following sections.

*3.2 Bubble definition*
*"I wonder how much it would take to buy a soap bubble, if there were only one in the world." Mark Twain.*

A bubble starts with a new opportunity or expectation. This can be a ground-breaking new technology, the access to a new market, or a significant technical trading event such as the breaking of a support line. In any case, there must be a good story about terrific future prospects. Smart money flows in at the early stage, which leads to a first wave of price appreciation. Attracted by the prospect of extrapolated higher returns, more investors follow. At some point, demand goes up as the price increases, and the price goes up as the demand increases. This is a positive feedback mechanism, which fuels a spiralling growth away from equilibrium. As when pulling the plug out of the bathtub, the equilibrium has been broken and there is no longer any serious price determination at the intersection of supply and demand.

This phenomenon should not be confused with the uncertainties concerning fundamental value under conditions of noise, as it was described by Fischer Black. During a bubble, the market has changed structurally and entered a completely new regime, which is entirely driven by sentiment and no longer reflects any real underlying value. At some point, often when liquidity starts to dry out due to central banks raising rates and/or foreign capital no longer flowing in, investors start questioning whether the process is sustainable. Often, this is followed in short order by a market collapse resulting from the synchronisation of sell orders. The crash occurs because the market has entered an unstable phase after a long maturation process associated with the inflation of the bubble. At this stage, like a ruler held up vertically on one finger, any small disturbance can trigger a fall. This mechanism is often not well understood and much controversy then arises about the cause of the crash. In reality, the cause is quite obvious and is found in the preceding years of exuberant bubble dynamics that have made the whole construct fragile. Thus, it is no surprise that it can collapse on the smallest of shocks. As we show below, this insight offers us the possibility of advanced diagnosis of developing bubbles, the Holy Grail for any investor interested in protecting his or her portfolio.

*3.3 Bubbles are processes of unsustainable growth*
*"Whenever you find yourself on the side of the majority, it is time to pause and reflect." Mark Twain.*

### 3.3.1 Cause and effect
By nature, a financial bubble is an unsustainable process in which the system is gradually pushed towards criticality. In a critical system, small events can have huge impacts. There is no point in arguing about local causes and consequences when the system has reached a critical state, as it is the criticality that matters.

This point is nicely illustrated by the many discussions and controversies on the cause of the recent global financial crisis. It is generally accepted that the 2007–2009 recession in the US was triggered, in some way, by the correction in the housing market after its peak early in 2006. As house prices started to level off, wealth extraction, associated with mortgage renegotiations and purchases made for pure capital gains, began to stall. In consequence, a growing number of homeowners found themselves owing more on their mortgages than their homes were worth. Consequently, the rate of defaults and foreclosures skyrocketed, especially in the subprime segment, with the particularly loose underwriting standards that had previously led to its disproportional growth. Additionally, the previous years had seen a wave of financial innovations. New structured credit products, like CDOs (collateralised debt obligations) and MBSs (mortgage-backed securities), were developed. Because of them, and the fact that many of them were additionally insured against losses by AIG, the international insurance and financial services company, the decline in house prices infected the global financial system, leading to the potential systemic risk of a global freeze.

This is, however, a quite unsatisfactory explanation, as the question of why exactly the US housing market stopped appreciating remains open. If it had continued to expand at its previous rate, fulfilling the expectation of most market participants based on its behaviour over the preceding century, in which real estate had almost never disappointed, it is likely that the crisis would have been averted, or at least postponed. In the following sections, we explain that house prices had followed an unsustainable track, which brought the market to a critical state characterised by the existence of an intrinsic end-point. Thus, growth could not continue and once it reached the critical point, anything might have triggered the change of regime.

It is now time to explain how unsustainable growth differs from sustainable growth and what precisely characterises a critical system.

### 3.3.2 Faster-than-exponential growth
*"For unto every one that hath shall be given, and he shall have abundance: but from him that hath not shall be taken even that which he hath." Matthew 25:29, King James Version [8].*

Generally, capital grows exponentially. That is why, most often, market movements or economic growth are reported in percentage returns, and why we are used to hearing phrases like "last year, GDP grew 1.7 percent", or, "last year, the Swiss Market Index rose 18 percent". This is caused by the mechanism of compound interest rates, which in science is called a proportional growth process.

A very good and clear definition can be found on Wikipedia [9]:
*"A proportional growth process is any of a class of processes in which some quantity, typically some form of wealth or credit, is distributed among a number of individuals or objects according to how much they already have, so that those who are already wealthy receive more than those who are not."*

In sociology, this is also called the "Matthew effect", in reference to the quotation above. The process is also often summarised in the aphorism *"the rich get richer and the poor get poorer"*.

When resources are unlimited, exponential growth can go on indefinitely. This is different in systems of finite size, where there is competition for limited resources, such as a species population growing under Darwinian competition. If a pair of sheep is put on an island, to begin with growth will be exponential, as the resources will be practicality unlimited for the small population. However, the available food can feed only a limited number of sheep. As the population approaches its limit, given by the carrying capacity of the island, the growth rate slows down due to the negative feedback of competition for the scarce food. Eventually, growth will stop and a dynamic equilibrium between the available food and the sheep population will be reached. Such a growth process under competition is called logistic growth and the equilibrium is the general result of the presence of negative feedbacks, which have a stabilising effect. Precisely this scenario unfolded when sheep were introduced on the island of Tasmania around 1800 [10]. The point of this explanation is that neither exponential nor logistic growth processes contain, in and of themselves, any endogenous cause of instability. In the first case, growth simply continues forever, in the second, growth stops and a steady state is reached.

The key ingredient that sets off an unsustainable growth process, which is a prerequisite for a financial bubble, is positive feedback. This is also called pro-cyclicality in economics, and is the exact opposite of negative feedback. Positive feedback is often caused by imitation: when investors display herd behaviour, a price increase triggers even greater demand due to the strengthening of the herd, and the equilibrium of supply and demand breaks down, as when pulling the plug out of the bathtub.

A classical example of a positive feedback mechanism that sets an unsustainable process in motion is the formation of droplets from a dripping tap. This is illustrated in Figure 1. Initially, the surface tension of the liquid keeps the droplet connected to the tap. But the droplet is pulled down by gravity and this gradually changes its shape. The process that leads to the fall is comparable to a climbing rope that snaps. When the first fibre breaks, the load is distributed over the remaining fibres. Each time a new fibre snaps, the load on the other fibres increases, leading ever more new fibres to break in what becomes a vicious circle. This is a positive feedback mechanism that leads to a runaway process: the failure of the rope. For the droplet, the thinner the neck holding it, the greater the stress applied to the liquid in this constriction, and the faster the flow that shrinks it; again a positive feedback. The thinning of the neck accelerates and ends in fracture, a finite-time singularity in the jargon of mathematicians. This translates into disconnection and the fall of the droplet.

In order to develop an operational financial-bubble model, we need to translate this insight into mathematical vernacular. There are no positive feedback mechanisms in the standard financial models, which assume that the growth of asset prices is essentially a stochastic proportional process, fuelled by the mechanism of compound returns or interest rates. This means that, apart from its volatility, a stock price is supposed to grow, on average, exponentially at a constant rate of return. When positive feedback is involved, the dynamics change drastically. Now, the growth rate is no longer constant, but starts growing itself, which makes the price follow a (faster-than-exponential) hyperbolic course until, at some point, the growth rate becomes so large that the price hits a wall and the model breaks down. In physics and mathematics, such a point is called a singularity. Figure 2 illustrates the difference between exponential and hyperbolic growth. It shows that the latter can be quite deceptive, following a gentler slope than the former for the major part of its development and then surpassing its growth rate in the final stage. Furthermore, exponential growth can continue mathematically ad infinitum, whereas hyperbolic growth reaches a point of instability at which the price process ceases to exist. This is the point where the system becomes critical. Using our mechanical metaphor, it is when the thin neck holding the droplet fractures or the climbing rope fails.

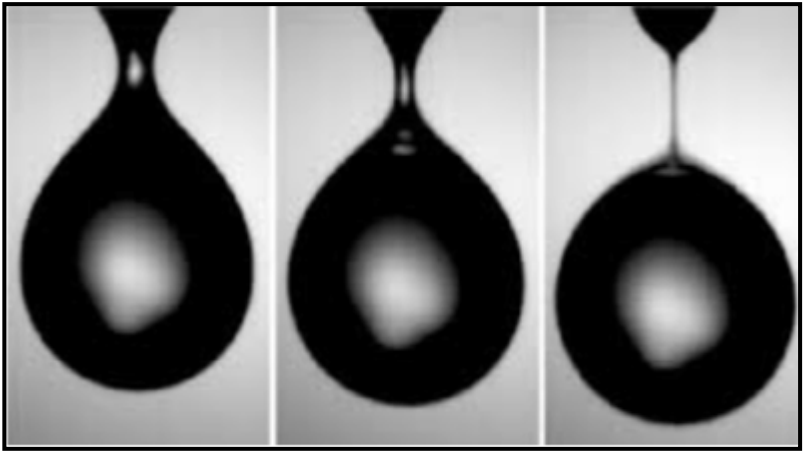

*Figure 1:* *The formation of a droplet is an example of a positive feedback mechanism that sets in motion an unsustainable process ending with a finite-time singularity, namely the fracture of the thin neck and the fall of the droplet (reproduced from [11]).*

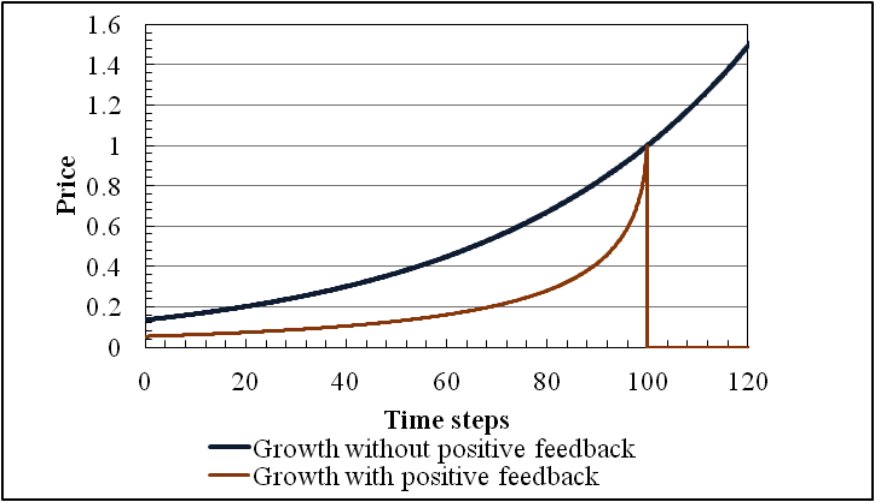

*Figure 2:* *The difference between exponential and hyperbolic growth. Growth without positive feedback can continue ad infinitum; growth with positive feedback diverges and reaches a critical point where the model breaks down. In this example, the singularity is put at time step 100. Because the model does not give any solution beyond that point, the value is then put at zero. Note also that, for the major part of the process, hyperbolic growth follows a gentler slope than exponential growth, overtaking its growth rate only towards the end.*

This is a very powerful insight that can be used to analyse financial time series. In principle, by verifying mathematically whether the price follows a hyperbolic course instead of an exponential one, it can be determined whether a positive feedback mechanism, which characterises the bubble phase of an asset, is at work. If this is the case, the price trajectory cannot be sustainable and a critical point will be reached at which a change in the market will occur. That is the point where the risk of a crash or a major correction is highest.

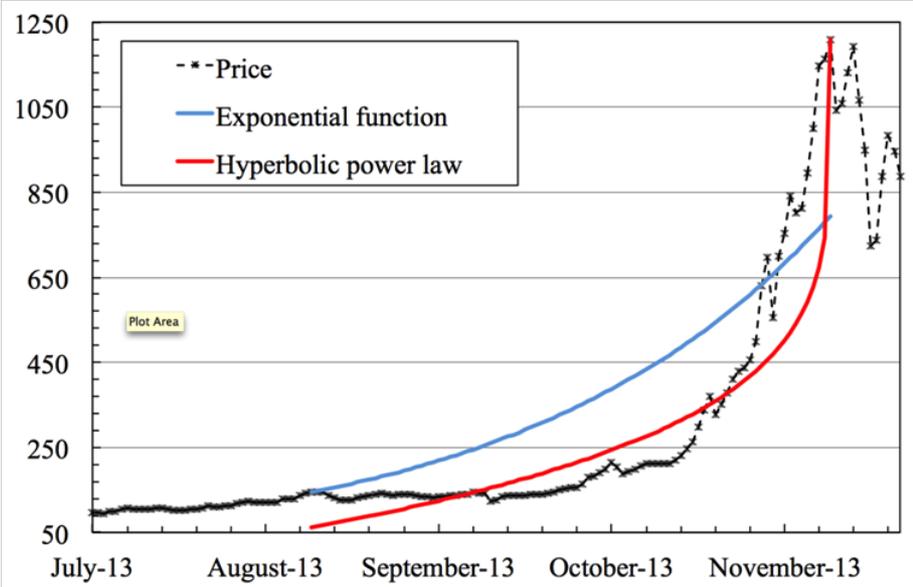

*Figure 3:* The price of Bitcoin before the major correction at the end of November 2013. A hyperbolic process is needed to reflect the steep rise in the price at the final stage of the bubble. This is the hallmark of a positive feedback mechanism and of the existence of a bubble
(Source: http://bitcoincharts.com/charts/mtgoxUSD#rg180ztgSzm1g10zm2g25zv).

The non-sustainable character of a price trajectory under the influence of positive feedback is illustrated in Figure 3, which shows the development of the price of Bitcoin (the digital currency introduced in 2009) before the major correction at the end of November 2013. It can be clearly seen that the exponential function (in blue) is not a good model for capturing the price dynamics, which is characterised by a slow initial rise followed by ever accelerating growth until the breakpoint. It attempts to portray the real dynamics with an average annualised growth rate of almost 700 percent, which is much too large for most of the period and too small for the final stage. The reason for this discrepancy is that the growth rate has not been constant but has itself been growing, as shown in Figure 4. This acceleration in the growth rate is captured well by the hyperbolic growth model. This analysis enabled the diagnosis that Bitcoin was in a bubble and that a correction would come when the market entered a critical state as a result of positive feedback mechanisms.

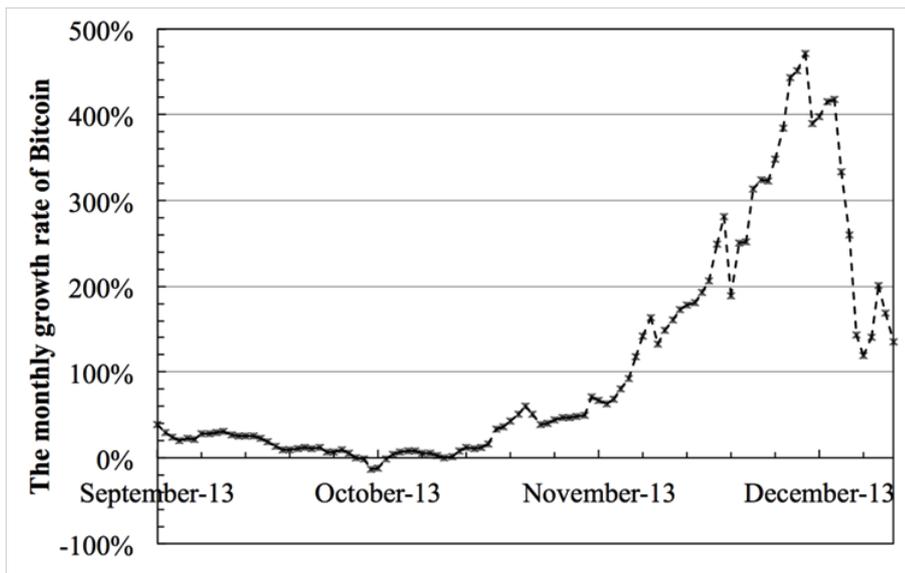

*Figure 4: The monthly growth rate of Bitcoin before the major correction at the end of November 2013. It can be clearly seen that the growth rate is itself growing; this is caused by a positive feedback mechanism, for instance herd behaviour. During this period of excessive growth of the growth rate itself, the asset is in a bubble phase (Source: http://bitcoincharts.com/charts/mtgoxUSD#rg180ztgSzm1g10zm2g25zv).*

There are many positive feedback mechanisms. They can be classified into two broad classes: (i) technical/rational and (ii) behavioural.

In the first class we find, for instance, option hedging and portfolio insurance techniques. Option hedging is the simple implementation of the Black-Scholes option theory, in which the risk associated with selling, say, a call option is eliminated (in a Gaussian world) by buying more of the underlying asset if the price has gone up recently and vice-versa. This is clearly a positive feedback strategy since the higher the price, the higher the demand becomes. Ronnie Sircar and George Papanicolaou have shown, for instance, that taking such hedging strategies into account, actually leads to a significant increase in the volatility of the underlying asset of the option [12].

Portfolio insurance techniques refer to the synthetic put option strategy developed by Leland and Rubinstein to hedge against market downturns. The problem is that its large-scale adoption in the 1980s by a number of big financial investment funds contributed significantly to the severity of the Black Monday crash on 19 October 1987, since it created strong selling pressure as the price went down. Trend-following investment strategies are clearly pro-cyclical, as, for instance, a recognised previous upward trend sends a signal to buy, supporting further price rises. We think that algorithm trading is also increasingly contributing to the creation of positive feedback periods on the markets, due to the adaptive and learning abilities of the algorithms, which tend to identify strategies with similar performance, leading to herd behaviour.

The crisis that started in the USA in 2007 has once again illuminated the pro-cyclical nature of the way banks finance companies by making more credit available in booms and by limiting credit in periods of contraction. Additionally, it confirmed the cyclical behaviour of deregulation in good times and reintroducing new regulation after crises. The learning process in business networks and the build-up of human capital are other positive feedback mechanisms. Different kinds of execution processes, such as stop-loss orders, market makers' bid-ask spread optimisation in response to past volatility, and the optimisation of trade-execution, also promote positive feedback loops on asset prices.

Positive-feedback behavioural mechanisms include imitation and herd behaviour. It can, in fact, be proved that imitation turns out to be the optimum investment strategy in conditions of serious uncertainty, since the price, which results from the aggregate polling of decisions, can be proxied by the aggregate sentiments of one's "friends". We see that bubbles often develop in the presence of exciting stories concerning a novel investment opportunity, which is actually characterised by extreme uncertainty concerning the real future outcomes. Imitation and herd behaviour then both fuel the bubble and become the best strategy – until the bubble bursts. Robert Cialdini has documented the "social proof" mechanism [13], in which people will do things that they see other people are doing. For example, if one person looks up at the sky, bystanders will then also look up to see what he is looking at. This apparent mindless behaviour may actually reflect our innate tendency to use simple clues and heuristics that offered better chances of survival at the time we evolved into hunter-gatherers about 200,000 years ago.

3.3.3 Singularity
We mentioned above the concept of a singularity, borrowed from physics and mathematics. This is a point at which a model breaks down and the equations no longer have any solution.

| Time | Population | Growth rate | Doubling time (years) |
|---:|---:|---:|---:|
| 0.00 | 2 | 2% | 34.65 |
| 34.65 | 4 | 4% | 34.65/2= 17.33 |
| 51.98 | 8 | 8% | 34.65/4= 8.66 |
| 60.64 | 16 | 16% | 34.65/8= 4.33 |
| 64.97 | 32 | 32% | 34.65/16= 2.17 |
| 67.13 | 64 | 64% | 34.65/32= 1.08 |
| 68.22 | 128 | 128% | 34.65/64= 0.54 |
| 68.76 | 256 | 256% | 34.65/128= 0.27 |
| 69.03 | 512 | 512% | 34.65/256= 0.14 |
| 69.16 | 1024 | 1024% | 34.65/512= 0.07 |

*Table 1: The first ten steps of a process in which an initial population of two grows at an initial rate of 2 percent and the growth rate doubles suddenly each time the population doubles, forming a succession of iterations.*

Take, for example, a population of two that increases with an initial growth rate of 2 percent. After 34.7 years, or roughly one generation, this population will have doubled. Now, suppose that the growth rate itself doubles with every new generation. Then, in the next step, the growth rate will be 4 percent and the doubling time 34.65/2=17.3 years; in the following step, the growth rate will be 8 percent and the doubling time 34.65/4=8.7 years and so on; Table 1 shows the first ten steps of this accelerating growth dynamic. As the process develops, the population grows without limit, reaching infinity in the finite time of 34.65 + 34.65/2+34.65/4 + 34.65/8 +... = 34.65 x (1+1/2 +1/4 +1/8+ ...). The infinite series illustrates Zeno's famous paradox, which bothered the classical Greek mathematicians and philosophers. Zeno argued that an arrow should never reach its target since at any time it has to cover half the remaining distance, and then again half the next remaining distance and so on, giving for the total remaining distance the infinite series 1/2 +1/4 +1/8+... . The Ancient Greeks struggled with the notion that such a summation containing an infinite number of summands could give a finite result. Starting with Cauchy's theory of convergence, we now know that 1+1/2 +1/4 +1/8+... is simply equal to 2. Thus, the process in Table 1 illustrates a finite-time singularity (see Figure 5), as the population goes through an infinite number of doublings in just 2x 34.65=69.3 years. Mathematically, the process cannot continue beyond that time, and change is unavoidable: this may be a crash or a substantial correction.

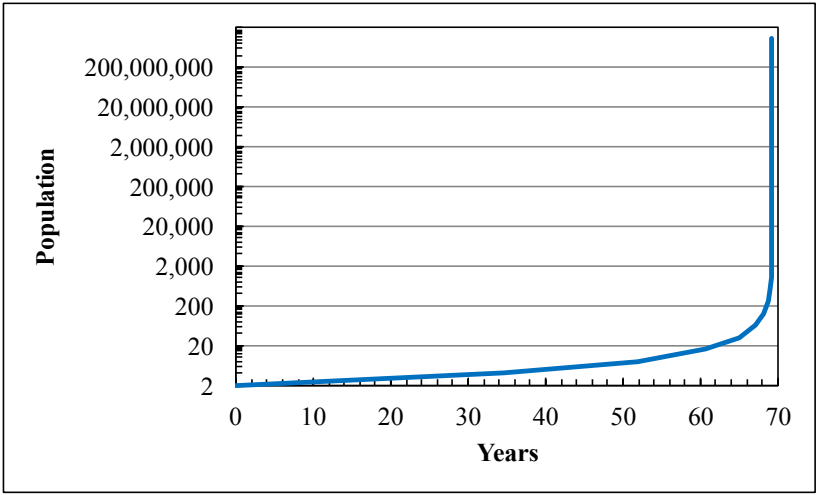

*Figure 5:* *When the growth rate itself grows, the process becomes unsustainable and ceases to exist at a critical point. Here, an initial population of two grows at an initial annual rate of 2 percent. Then, we assume that the growth rate doubles at each population doubling. The process reaches a finite-time singularity and cannot continue beyond 69.3 years. Mathematically, there must be a change at that point; in real life the change will occur slightly before it. Notice that the y-axis has a logarithmic scale. In such a representation, an exponential process with a constant growth rate would show as a straight line whose slope is the growth rate.*

Interestingly, in economics and finance, models that contain a finite-time singularity are automatically discarded because they violate the principle of "proof of existence" required for the construction of a bona-fide theory. In general, economic models are designed so as to ensure the existence of an equilibrium, or at least a solution at any point in time. Here, we take another view by accepting that the market can have different phases with distinct dynamics, depending on whether the pricing mechanism follows a sustainable or a non-sustainable process. As a matter of fact, the non-existence of a solution is the key point in our methodology for predicting the end of a bubble. Rather than making naïve forecasts based on extrapolated trends, our intrinsically nonlinear approach allows us to identify different market regimes. Most importantly, it characterises the end of unsustainable phases, when the risk of a crash or a correction is greatest; the fact that the model actually breaks down contains the really important information.

Such finite-time singularities may not be part of an economist's mathematical toolbox yet, but they play an important role in natural sciences. Equations in physical theories often describe the emergence of singularities occurring in finite time, which are associated with important physical phenomena. For example, the Theory of General Relativity, developed by Einstein predicts that a black hole forms when a very massive star collapses at the end of its life cycle, and that a gravitational singularity is created in finite time where space-time curvature becomes infinite. Other examples may be found in the theory of planet formation in the solar system, turbulence in fluids or plasmas, front genesis in meteorology, rupture and material failure, earthquakes, surface instabilities causing spikes or jets, and the Euler disc.

### 3.3.4 US house prices revisited

We now have all the tools to revisit, and answer, our initial question. If a slow-down and levelling off of US house prices was the factor that triggered the recent financial crisis, what made house prices stop rising? Figure 6 shows the levels and annual growth rates of US real house prices. Three corrections can be observed: in 1979, 1989 and the massive fall starting in 2006 that set off the crisis in 2007. It can be clearly seen that each drop was preceded by a steep rise in the growth rate. In each case, then, real-estate prices were on an unsustainable faster-than-exponential trajectory, which necessarily had to end. As we have explained above, the bursting of the bubble was the inevitable consequence of hyperbolic, faster-than-exponential price growth. In fact, using these concepts and the technical implementation described below, in June 2005 we published in the international scientific digital archive (http://arxiv.org/abs/physics/0506027) a warning that the US real-estate market would experience radical change by mid-2006. The article was published in a scientific journal afterwards in 2006 [14]. This is a clear illustration of how the market functions like a dynamic system that experiences phases of unsustainable growth followed by corrections, and is in no way a system in equilibrium in which the fundamental price is discovered at the intersection of supply and demand.

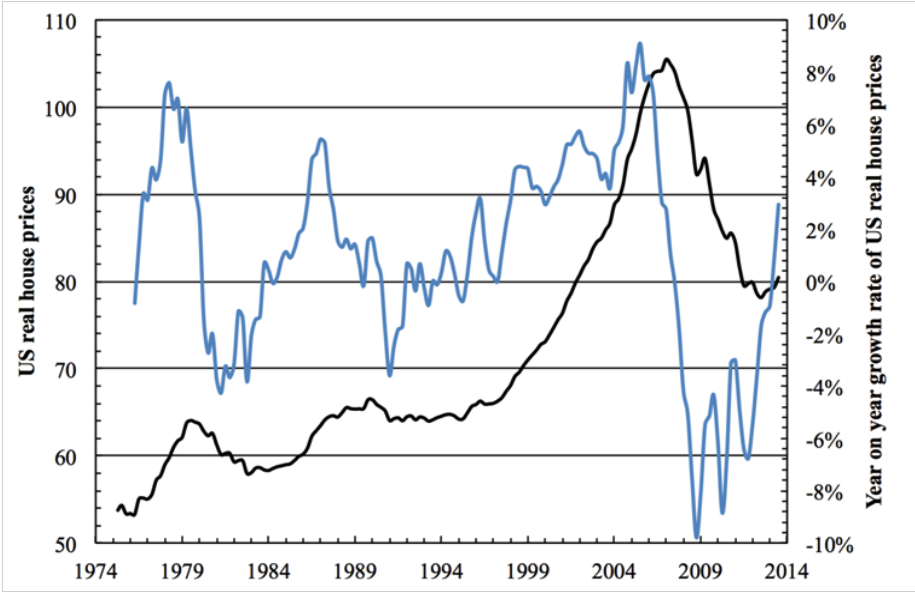

*Figure 6:* US real house prices between 1974 and 2014. Levels are shown in black and should be read on the left-hand axis. Annual growth rates are shown in blue and should be read on the right-hand axis. Three peaks in the growth rate are immediately followed by a correction. When the growth rate itself grows, the process becomes unstable and a correction follows around the critical point embedded in the faster-than-exponential growth process. (Source: Federal Reserve Bank of Dallas international house price dataset, http://www.dallasfed.org/institute/houseprice/)

*3.4 Market structure*
*"The universe is built on a plan the profound symmetry of which is somehow present in the inner structure of our intellect." Paul Valery.*

We have argued that a bubble is essentially an unsustainable process that is fed by a positive feedback mechanism. As a consequence, the growth rate itself grows and the price follows a hyperbolic power law trajectory ending in some critical point or singularity where the risk of a crash or a correction is very high. An example of such an idealized process was shown in figure 5.

In reality, the process will not always be so smooth. Often, typical patterns of oscillations can be observed that are caused by specific dynamic and structural features of the market. By combining the hyperbolic power law that was described in the previous section and a model for these distinct oscillations that will be explained here, we will get to the log-periodic power law (LPPL) model, which is the workhorse of our present bubble detection methodology.

In a ground-breaking paper in 1998 [15], Robin Dunbar, a professor of Evolutionary Psychology and Behavioral Ecology at the University of Liverpool, presented his "social brain hypothesis". He argued that the evolution of primate brains was driven by the need to function in increasingly larger social groups. Dunbar based his conclusion on the observation of a strong relationship between the social group size and the relative neocortex volume in primates. Interestingly, his model revealed that, based on the brain size, humans have a cognitive limit of approximately 150 individuals with whom a coherent personal relationship can be maintained.

More recent research by Zhou et al. [16] has refined and extended this hypothesis by discovering that this social group of 150 individuals has a marked discrete internal structure comprising a hierarchy of sub-groupings. In this order, the core is called the support clique. This is the nucleus, or family of three to five persons that support each other during periods of severe emotional or financial distress. The next level on top of this is called the sympathy group. It contains the persons with whom we have special ties, like, for example, the people we see at least once per month. This unit includes about twelve to twenty individuals, enclosing three to four support groups. The following level in this pattern is the band. That is a party of about 30 to 50 persons. Ethnographic data on hunter-gatherer societies has shown that this group corresponds to the typical size of overnight camps. This internal organization in support cliques, sympathy groups and bands is dynamical and may change continuously over time, but their membership is always drawn from the same set of typically 150 individuals, defining in total 5 levels in the hierarchical organization when including the ego.

The comprehensive and systematic study of Zhou et al [16] used a large data set on human grouping patterns documented by many previous scholars. By applying fractal analysis, they identified, with high statistical confidence, a discrete hierarchy of group sizes with a preferred scaling ratio between three and four. They write: *"… rather than a single or a continuous spectrum of group sizes, humans spontaneously form groups of preferred sizes organized in a geometric series approximating 3-5, 9-15, 30-45, etc. Such discrete scale invariance could* *be related to that identified in signatures of herding behavior in financial markets and might reflect a hierarchical processing of social nearness by human brains."*

Interestingly, a similar organizational structure with a comparable scaling ratio of three to four can be found in the military where one typically finds sections (or squads), platoons, companies, battalions, regiments, divisions and corps. Zhou et al. [16] ask the following natural question: *"Could it be that the army's structures have evolved to mimic the natural hierarchical groupings of everyday social structures, thereby optimizing the cognitive processing of within-group interactions?"*

The point of this parenthesis into the study of social structures is that, if traders and investors organize in the same natural hierarchy, it will be reflected in the way they imitate each other and as such have an impact on their coordinated buy and sell orders. As a consequence, this discrete social structure will have a real effect on the pricing mechanism during phases of strong herding. Mathematically, this insight translates into the creation of a specific pattern in the price, with oscillations that are accelerating and form a discrete temporal hierarchy mirroring somehow the discrete social hierarchy of traders. Mathematically, the underlying symmetry is called discrete scale invariance (DSI) and is the manifestation of the hierarchical (non-continuous) scaling in the organization of traders and investors. As a result, during bubble phases, ever increasing oscillations with decreasing amplitude will be seen in the price. This is called log-periodicity.

Remarkably, log-periodicity and DSI may occur spontaneously for entirely different causes. It may emerge from a purely dynamical origin, without a pre-existing hierarchy, as a consequence of the interplay between value traders and chartists, also called trend followers or technical traders. Suppose that an asset is quoted significantly under its expected fundamental value. In that case, buy orders by value traders will push the price up. This will generate a momentum that will be picked up by trend followers. As a consequence, the price will overshoot the fundamental value due to the inertia of investors when reassessing their expectation in the presence of uncertainty. The value traders will start selling and this will now drive the price down. A new momentum, this time in the opposite direction will be detected and be picked up again by the trend followers. This will cause the price to undershoot the fundamental value, what will prompt value traders to start buying again, setting in motion a new cycle. The interaction between technical traders that push the price out of balance and value traders that introduce a restoring force will generate oscillations. Taking into account that, in practice, traders make decisions to buy or sell when their signals are sufficiently strong, this creates a nonlinear threshold-like process and the oscillations become nonlinear. Nonlinear oscillations have, by definition, a frequency that depends on the amplitude. During the fast price appreciation of a bubble phase, the oscillations will have an ever-increasing frequency, implying that the inertia of investors when reassessing their expectation decreases.

Such log-periodic oscillations have been reported in many dynamical systems like hydrodynamic turbulence, chemical and biological growth processes, material rupture, earthquakes and financial crashes [17]. These may have different types of self-similar or fractal-like properties and may hold pockets of predictability of an imminent regime shift. An example from every-day life can be found in traffic jams where often oscillations can be experienced before the "phase transition" between the free-flowing and congested states. Additionally, it turns out that bats and dolphins use log-periodic chirps for echo-location during navigation and foraging. This is not incidental as such log-periodic chirps provide the optimum resolution of obstacles at multiple scales [18].

*3.5 Diagnosing bubbles*

Our bubble model combines these specific oscillations with the hyperbolic power law in one single equation called the log-periodic power law (LPPL). The term "log-periodic" (LP) refers to the part of the equation that models the accelerating oscillations and the term "power law" (PL) handles the faster-than-exponential rise in the price, the fact that the growth rate itself grows due to feedback.

The exact formula is given in Figure 7. Three different components are joined together to make one single model. The first term is the plain mathematical description of the smooth hyperbolic power law. The second term, controlling the amplitude, and the third cosine term combine to form the log-periodic oscillations.

$$\ln(P) = \underbrace{A + B\,(t_c - t)^m}_{1} + \underbrace{C\,(t_c - t)^m}_{2} \cos\left(\underbrace{\omega \ln[t_c - t] - \phi}_{3}\right)$$

*Figure 7: The log-periodic power law (LPPL). In one single model, there are three precursors of the same critical time $t_c$ at which or around which a change of regime will occur. This equation has seven parameters. The constant term A is the expected value of the log-price at the peak when the end of the bubble is reached at time $t_c$, the critical time. B and C respectively control the amplitude of the power law acceleration and of the log-periodic oscillations. The exponent m, usually between 0 and 1 for a bubble, quantifies the degree of super-exponential growth. The log-periodic angular frequency $\omega$ is related to the preferred scaling ratio of the temporal hierarchy of oscillations converging to the critical time $t_c$. Finally, $\phi$ is a phase that quantifies the time scale of the oscillations.*

Each of these three parts models a separate process that in itself develops towards the same critical time $t_c$:
1. For 0<m<1, the first part of the equation takes care of the positive feedback mechanism, when price growth becomes unsustainable, and at $t_c$ the growth rate becomes infinite;
2. The second part of the formula causes the amplitude of the oscillations to drop to zero at the critical time $t_c$;
3. Part three models the frequency of the oscillations, which becomes infinite at $t_c$.

Hunting for bubbles essentially involves looking for this specific pattern in the data. If the pattern of the log-periodic power law can be found with sufficiently strong statistical confidence, this is a clear indication of an unstable process, a bubble. This is illustrated in Figure 8, in which a positive bubble pattern can be seen in the S&P500 in the years before, and a negative bubble pattern in the period during the crisis.

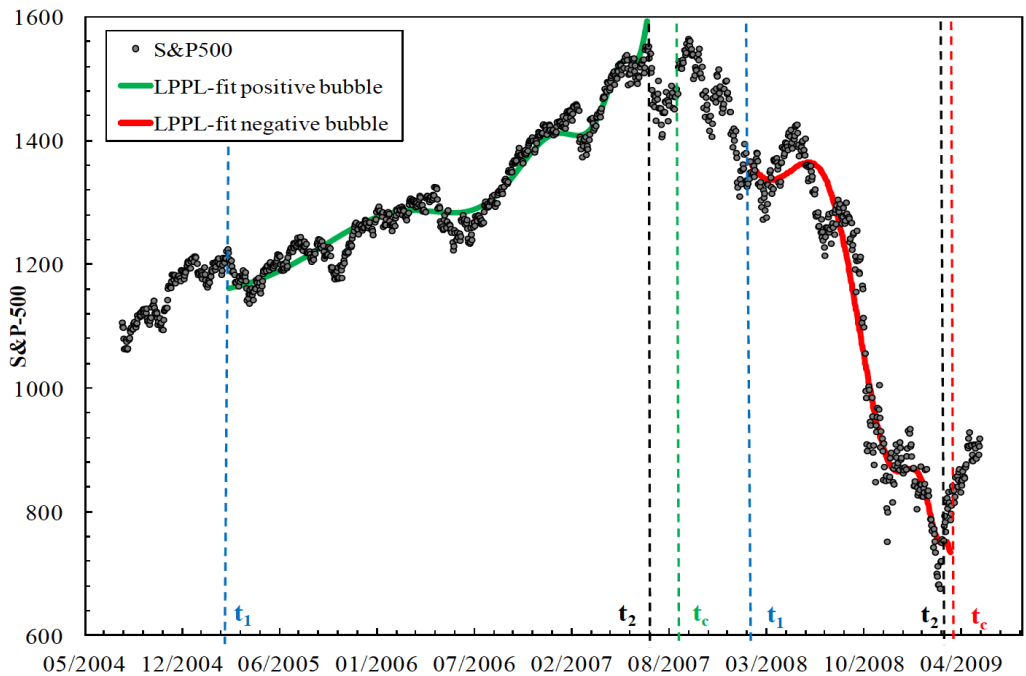

*Figure 8: The pattern of a positive bubble (in green) and a negative bubble (in red) can be seen in the S&P500 before and during the credit crisis. Notice the distinctive pattern combining oscillations with a faster-than-exponential rise (or drop for a negative bubble) in the price. For both bubbles, the LPPL model is fitted to the data in the window between $t_1$ and $t_2$. The critical time $t_c$ is determined from the calibration of the model to the data in the corresponding windows; $t_c$ is the most probable time for the change in regime to occur, which may be a crash after a positive bubble or a rally following a negative bubble.*

Calibrating this model and its siblings developed in our group is not easy [19]. Extreme care should be exercised to prevent over-fitting (when a good apparent fit is the result of mere chance) and obtain robust calibrations. Here, we can barely touch on this vast field, simply stating that, after almost two decades of experimenting and improving the methodology, we have developed a process that is open to systematic ex-ante real-time testing, which we began in 2009 with the financial bubble experiments in the financial crisis observatory at ETH Zurich. Interested readers are encouraged to review the resources provided on the website of the Financial Crisis Observatory [20] and check the publications that report the progress and tests [21].

Our research suggests that bubbles occur on all possible time scales. This means that they can form, develop and mature over a period of an hour, but also over a period of a century. As we do not know the time base of the bubble, the key to hunting for bubbles is to scan the data using different window sizes. As a result, during bubbles, we will not have a single calibration with one critical time, as is indicated in Figure 8 for the sake of simplicity, but a full ensemble. This allows us to carry out a statistical analysis as shown in Figure 9. This shows the result of an analysis of the CHF/EUR exchange rate in the turbulent times when it flirted with parity in response to the European crisis. The alarm index for diagnostic of the bubble, calculated from the ensemble of fits, is shown in green; the black line represents the point at which there was a 50 percent probability of a crash; the pink area gives the broader time frame within which the correction is to be expected. These analyses were performed ex-ante, before the "crash" of the Swiss franc actually occurred, as part of research performed in the FCO at ETH Zurich. Another example is given in Figure 10. This shows the prediction of a major correction that occurred after the build-up of a massive bubble for oil and other commodities in the middle of 2008.

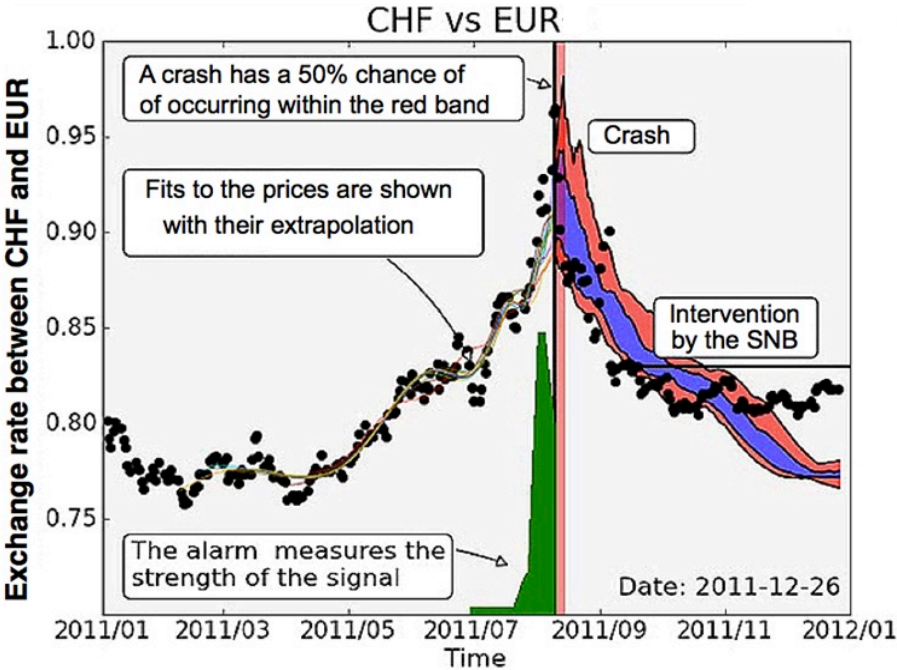

*Figure 9: The CHF/EUR exchange rate in 2011 when the exchange rate flirted with parity. The figure shows the results of a statistical analysis of an ensemble of bubble model fits. The ensemble is obtained by scanning the data using different window sizes and combining them in a statistical analysis.*

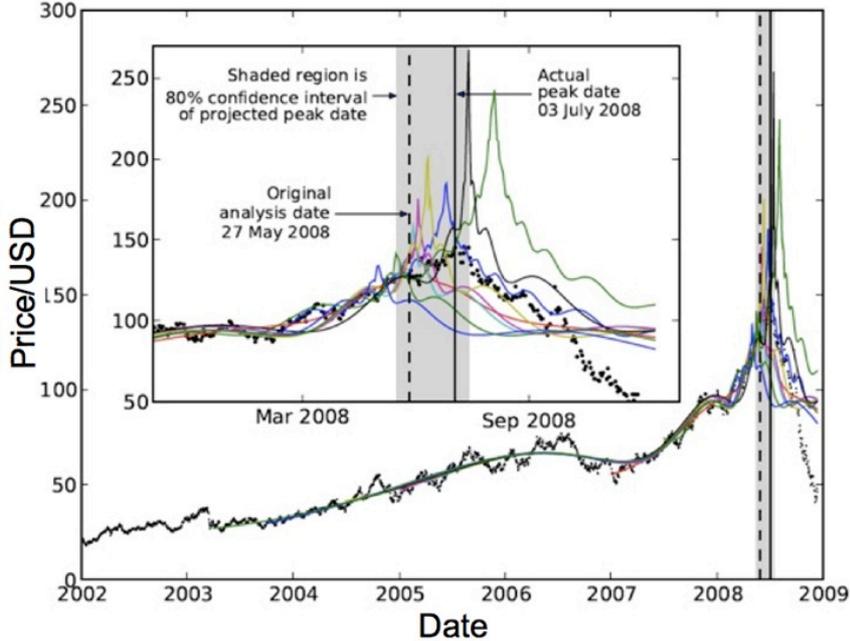

*Figure 10: The historical development of oil prices (dotted line). The vertical dashed line shows the last observation used to calibrate our model. The coloured curves show different LPPL fits. The inset shows a magnification around the time of the peak. The correction was expected to occur with an 80 percent probability in the grey shaded zone. This indeed happened (reproduced from [22]).*

As can be seen in the figures, both forecasts, which were made before the correction actually happened, turned out to be highly accurate. Over the past fifteen years, our group has performed a whole set of ex-ante predictions. For a detailed overview, we refer to one of our previous papers [4], the book *Why Stock Markets Crash* [23] and, again, the website of the Financial Crisis Observatory [20].

3.*6 Distinction between the notion of a positive/negative bubble and over-/undervaluation*
It is important to stress that our bubble identification model is based on relative price changes and on the characterisation of the full dynamics, which are compared to a benchmark of exponential growth. This is fundamentally different from the more conventional approach of taking a snapshot of the price of an asset at a given time and comparing it with the best estimate of its fundamental value. In that approach, a bubble is identified when the observed market price is significantly higher than the estimated real value. In our methodology, rather than focusing on individual snapshots, we take a dynamical, movie-like view, in which it is not so much the price level that counts but the way it was reached. In effect, our methodology has nothing to say in itself about any possible mispricing. The key concept is that we identify dynamic market situations that are not sustainable and are bound to collapse at the climax of a positive bubble, or rally in the aftermath of a negative bubble. The collapse (or rally) may occur, again transiently, even if the price still remains too low (or high) compared to the fundamental value.

It is important to distinguish two notions: (i) the transient non-sustainable exuberant market situations that our methodology identifies, and (ii) the existence of profound mispricing. Often, the two are closely related, as in the case documented in Figure 9, showing the CHF/EUR exchange rate during the European crisis in 2011. In the years before the crisis, the value of the euro in Swiss francs hovered in a range between 1.5 and 1.7, and it could be argued that this range was a market proxy for the fundamental value. In this sense, the accelerated appreciation of the CHF reaching parity in August 2011, with its characteristic super-exponential pattern, and the conclusion of a coming change, in the form of a collapse, resonates nicely with the belief that its correct fundamental value was (and still is) probably much lower.

At other times, mispricing might proceed through a series of transient positive or negative bubbles, with several changes of regime. These two aspects actually differentiate two characteristics of bubble phenomenology: (a) the revelation, via market dynamics resulting from the collective actions of investors, of the existence of transient non-sustainable situations that will collapse or recover; (b), the possibility of very substantial, but not yet apparent, mispricing, indicating to smart investors the presence of an opportunity to sell or buy, perhaps over a period of many years, while the market has not yet collectively even acknowledged such a possibility. It is thus useful to distinguish between assets that are fundamentally over-/undervalued and assets that develop a positive -/negative bubble.

This discussion reveals how difficult the identification of bubbles is. Essentially, the academic literature is mired in the debate on how to calculate mispricing. And few indeed are the professionals who are consistently correct in their assessment of real value. In other words, as explained briefly in section 3.1, determining the fundamental value of an asset means entering a minefield of difficulties and assumptions; it absolutely needs to be done, but stable successes are rare. It is even doubtful whether a determination of fundamental value can be genuinely falsifiable in the Popperian sense. This stalemate has been one of our main motivations for developing an alternative approach to the identification of bubbles, in the form of a robust diagnostic of dynamic non-sustainable overvaluation.

**4- Synthesis**

After the thirty years of unsustainable debt growth and excessive financial expansion associated with a general belief in the illusory perpetual money machine [4], which collapsed abruptly in the 2007–2008 financial crisis, it might have been thought that the parties involved (regulators, policy-makers, banks and investors) would have learned from their mistakes, and taken measures to prevent the occurrence of new bubbles and other financial excesses. But it turns out that not much has been done, and the approach is still to push for growth through consumption, via debt, now sovereign rather than private, and with an even larger financial sector more concentrated in fewer huge banks that are more systemically important. It also seems that central bankers have the unofficial, but increasingly obvious, mandate to steer stock markets upward, somehow hoping that the artificial rise, which should normally discount positive future expectations on economic growth, becomes self-fulfilling. And clearly, since the rebound on the US stock market in March 2009 there has been a series of bubbles, relatively short-lived but nevertheless of substantial scale, and corresponding corrections.

The daily analysis of financial markets in our observatory at ETH Zurich offers a sobering assessment of the presence of numerous bubbles in the world. These are in general not as massive as those that expanded before, and led to, the 2008 crisis, simply because they are developing over shorter periods. It also seems that there is an acceleration in the rate of formation and demise of bubbles, probably as a result of increasing algorithmic trading, more global interconnections between assets via ETFs and other financial instruments. Moreover, the importance of politics is drastically changing the balance of investments, in particular due to the successive waves of quantitative easing and extraordinary low interest rates that act as catalysts for a distorted allocation of resources. Since 2009, we have seen funds chasing returns in a twisted zero-rate environment, being seduced by the hyping of short-lived opportunities, such as the emerging markets, which promised so much a few years ago and are now in full correction.

The models and the methodology that we have outlined in the previous sections are at the core of the Financial Crisis Observatory (FCO) of the ETH in Zurich. Using the Brutus supercomputer, twenty five thousand financial time series are monitored on a daily basis looking for that exact specific bubble fingerprint that we have explained here. This allows us to have a continuous overview of emerging bubbles in the global financial markets [25], which we synthesize in a coherent report. The report, available in the Notenstein white paper series (2014), with the title "Financial Bubbles: Mechanism, diagnostic and state of the world (Feb. 2014)", presents our view on the status of positive and negative bubbles in the financial markets, as of the end of January 2014 [26].

Our research gives strong quantitative support to the hypothesis developed by Hyman Minsky [24] that financial markets are inherently unstable, with speculative and Ponzi-type finance being the typical developments on unregulated or loosely regulated financial markets. While technological bubbles may have long-term benefits [2,3], the development of policies that create financial bubbles, and trigger an even greater need for investors to speculate, does not seem to be the right direction to be going in. Financial markets have essential functions in our economic system: pricing, providing efficient allocation of capital, and enabling the diversification of risk. But creating bubbles, riding them and trading in them, will not solve the problems that led to the crisis of 2007–2008. Also in financial markets, the law of diminishing returns applies. Hence "more market" does not necessarily mean "better market". In the face of the waves of bubbles that encourage tactical, dynamic investing styles, it is time for economic and financial policy-makers to prioritise and promote "slow investment" associated with real fundamental value, for instance by taxing short-term (algorithmic) trading so that financial markets can once again fulfil their function as catalysts of prosperity.